# Chargaff's "Grammar of Biology": New Fractal-like Rules

*Michel E. Beleza Yamagishi*[1]

Embrapa Informática Agropecuária
Applied Bioinformatics Laboratory
209, André Tosello Av.
Campinas-SP-Brazil

*Roberto H. Herai*

Universidade Estadual de Campinas
Biology Institute
Campinas-SP-Brazil

## *ABSTRACT*

*Chargaff once said that "I saw before me in dark contours the beginning of a grammar of Biology". In linguistics, "grammar" is the set of natural language rules, but we do not know for sure what Chargaff meant by "grammar" of Biology. Nevertheless, assuming the metaphor, Chargaff himself started a "grammar of Biology" discovering the so called Chargaff's rules. In this work, we further develop his grammar. Using new concepts, we were able to discovery new genomic rules that seem to be invariant across a large set of organisms, and show a fractal-like property, since no matter the scale, the same pattern is observed (self-similarity). We hope that these new invariant genomic rules may be used in different contexts since short read data bias detection to genome assembly quality assessment.*

[1] Corresponding author: michel@cnptia.embrapa.br

# *INTRODUCTION*

In 1944, Oswald T. Avery (1877-1955) published a remarkable paper where the final sentence reads as follows: "The evidence presented supports the belief that nucleic acid of the desoxyribose type is the fundamental unit of the transforming principle of Pneumococcus type III" (Avery, 1944). In the early 40s, the knowledge about that strange macromolecule called "nucleic acid of the desoxyribose type" was scanty. Avery was the first to realize that DNA played a major role in the process of inheritable alterations of a cell. Erwin Chargaff (1905-2002) was deeply impressed by this discovery, and expressed his feeling saying "for I saw before me in dark contours the beginning of a grammar of Biology" (Chargaff, 1971). We are not sure what he really meant by "grammar of Biology", but he took it seriously because afterwards he relinquished all he was doing, and started his own DNA research journey.

After the development of a method for the precise chemical characterization of nucleic acids, Chargaff, in 1950, observed, using current language, that in any double-stranded DNA segment , the Adenine, A, and Thymine, T, frequencies are equal, and so are the frequencies of Cytosine, C, and Guanine, G (Chargaff, 1950). This observation is known as Chargaff's first parity rule. Watson and Crick were acquainted with this rule, and used it to support their famous DNA double-helix structure model (Watson & Crick, 1953). Chargaff also perceived that the parity rule approximately holds in the single-stranded DNA segment. This last rule is known as Chargaff's second parity rule (CSPR), and although it is not well understood, it has been confirmed in several organisms (Mitchell & Bride, 2006).

As CSPR is consistent with the DNA double-helix model, much effort has been devoted to understand the second rule (Forsdyke & Bell, 2004). Originally, CSPR is meant to be valid only to mononucleotide frequencies. But, it occurs that oligonucleotide frequencies follow a generalized Chargaff's second parity rule (GCSPR) where the frequency of an oligonucleotide is approximately equal to its complement reverse oligonucleotide frequency (Prahbu, 1993). This is known in the literature as the *Symmetry Principle*. As correctly pointed out by Forsdyke, higher order equifrequency does imply lower order, and he conjectured that the original CSPR was actually a particular case of a higher order parity rule (Bell & Forsdyke, 1999). This hypothesis led us to try to extend our former mathematical model (Yamagishi & Shimabukuro, 2008) to oligonucleotide frequencies and of other organisms, since it was originally supposed to be valid only to the human genome and to mononucleotide frequencies. Unfortunately, this task was not straightforward, and it took almost four years to be completed. The main reason was that it was necessary to introduce new concepts, most of them mathematical in nature such as the math table that will soon be introduced.

In this work, we present new higher order frequency rules which remained unnoticed until mathematical concepts such as partition set was used to organize oligonucleotides into equivalence classes. To the best of our knowledge, these new rules show for the first time that oligonucleotide frequencies do have invariant properties across a large set of genomes, and these rules, regardless the number of nucleotides remains the same (self-similarity).

## *METHODS*

Some mathematical definitions are necessary. The first definition is the k-word dictionary. As we are dealing with finite genomic sequences our alphabet, denoted by $\mathcal{A}$, consists of four letters $\mathcal{A} = \{A, C, G, T\}$, therefore the k-word dictionary, denoted by $\mathcal{W}^k$, is defined as the set of all words with exactly k letters (nucleotides). The cardinality, i. e., the number of elements of $\mathcal{W}^k$, denoted by $|\mathcal{W}^k|$, is given by $|\mathcal{W}^k| = 4^k$.

For instance, if k=2, then $|\mathcal{W}^2| = 4^2 = 16$ and the dictionary is

$$\mathcal{W}^k = \{AA, AC, AG, AT, CA, CC, CG, CT, GA, GC, GG, GT, TA, TC, TG, TT\}$$

Actually, most part of the following definitions are not new. For instance, k-words have been called k-mers in a recent study of genome inverse symmetry (Kong et al., 2009). However, we decided to use k-words instead of k-mer in order to reinforce the Chargaff's linguistic analogy.

### *Operators over $\mathcal{W}^k$*

In Molecular Biology, it is usual to compute the complement reverse of a given DNA sequence. This operation may be decomposed into two independent operators: the reverse $\mathcal{R}$ and the complement $\mathcal{C}$ operators. Although their definitions are well known, we define them below for completeness.

**Definition 1**: Given any k-word $w \in \mathcal{W}^k$, the reverse of w, denoted by $\mathcal{R}$ is also a k-word in $\mathcal{W}^k$ whose letters are w letters in the reverse order. Let $w = a_1 a_2 \ldots a_{k-1} a_k$, then $\mathcal{R}(w) = a_k a_{k-1} \ldots a_2 a_1$, where $a_i \in \mathcal{A}, i = 1, \ldots, k$.

**Example 1:** Let $w = ATGT. \mathcal{R}(w) = TGTA$.

Definition 2: Given any k-word $w \in \mathcal{W}^k$, the complement of w, denoted by $\mathcal{C}$, is also a k-word in $\mathcal{W}^k$ whose letters are the complement of w. Let $w = a_1 a_2 \ldots a_{k-1} a_k$, then $\mathcal{C}(w) = \mathcal{C}(a_1)\mathcal{C}(a_2) \ldots \mathcal{C}(a_{k-1})\mathcal{C}(a_k)$, where $\mathcal{C}(A) = T, \mathcal{C}(T) = A, \mathcal{C}(C) = G, \mathcal{C}(G) = C$ and $a_i \in \mathcal{A}, i = 1, \ldots, k$.

**Example 2:** Let $w = ATGT. \mathcal{C}(w) = TACA$.

Both operators $\mathcal{R}$ and $\mathcal{C}$ are involutions.

**Definition 3:** An operator is an involution when its inverse operator is the operator itself. Mathematically, $\mathcal{C}(\mathcal{C}(w)) = w$ and $\mathcal{R}(\mathcal{R}(w)) = w$.

**Example 3:** Let $w = ATC. \mathcal{C}(\mathcal{C}(ATC)) = ATC$, and $\mathcal{R}(\mathcal{R}(ATC)) = ATC$.

The complement reverse operator is defined as the composite operator $\mathcal{C} \circ \mathcal{R} = \mathcal{C}(\mathcal{R}(w))$ or $\mathcal{R} \circ \mathcal{C} = \mathcal{R}(\mathcal{C}(w))$. The definitions of $\mathcal{R}(w)$ and $\mathcal{C}(w)$ imply that $\mathcal{R}(\mathcal{C}(w)) = \mathcal{C}(\mathcal{R}(w))$, therefore we will use both forms interchangeably.

## *Induced k-word Set Partition*

The operators $\mathcal{R}$ and $\mathcal{C}$ and their compositions induce a set partition over $\mathcal{W}^k$.

**Definition 4**: A partition of a set $\mathcal{X}$ is the decomposition of $\mathcal{X}$ into disjunct and non-empty subsets.

Example 4: Let $\mathcal{X} = \{AA, AC, AG, AT, CA, CC, CG, CT, GA, GC. GG, GT, TA, TC, TG, TT\}$. A set partition of $\mathcal{X}$ is $\mathcal{P} = \{\{AA, AC, AG, AT\}, \{CA, CC, CG, CT\}, \{GA, GC, GG, GT\}, \{TA, TC, TG, TT\}\}$. This particular set partition decomposes $\mathcal{X}$ in four disjunct (the intersection of any two subsets is empty) and non-empty subsets.

An example of set partition, when k=3, is the codon table used by organisms to translate codons (three consecutive nucleotides) into amino acids.

The number of partitions of a set $\mathcal{X}$ depends on the number of elements in $\mathcal{X}$, and it is given by the Bell number. An n-element set $\mathcal{X}$ has $\mathcal{B}_n$ partitions given by the recursive formula:

$$\mathcal{B}_n = \sum_{k=0}^{n} \binom{n}{k} \mathcal{B}_k$$

where $\mathcal{B}_0 = \mathcal{B}_1 = 1$. The first 10 Bell numbers are: 1, 1, 2, 5, 15, 52, 203, 877, 4140, 21147. In Example 4 the set $\mathcal{X}$ has 16 elements, therefore it has $\mathcal{B}_{16} = 10.480.142.147$ possible set partitions, i. e., the number of set partitions of a 16-element set is about $10^{10}$ which is a huge number.

Therefore, the operators $\mathcal{R}$ and $\mathcal{C}$ and their compositions induce a very particular set partition over $\mathcal{W}^k$ defined as

**Definition 5:** We say that two words $w_i, w_j \in \mathcal{W}^k$ belong to the same subset or the same *Equivalence Class*, if, and only if, one of them is obtained from the other through the operators $\mathcal{R}$ and $\mathcal{C}$ or any of their compositions.

**Example 5:** If k=1, then the set partition induced by operators $\mathcal{R}$ and $\mathcal{C}$ and their compositions is

$$\mathcal{P}^1 = \{\{A,T\},\{C,G\}\}$$

Observe that A and T belong to the same equivalence class because $\mathcal{C}(A) = T$ and $\mathcal{C}(T) = A$, the same can be said about the equivalence class elements C and G. The $\mathcal{R}(w)$ operator is not necessary when k=1.

Example 6: If k=2, then the 2-word set partition is

$$\mathcal{P}^2 = \{\{AA,TT\},\{AT,TA\},\{CC,GG\},\{CG,GC\},\{AC,CA,TG,GT\},\{AG,GA,CT,TC\}\}$$

In this example, it is clear that $\mathcal{R}$, $\mathcal{C}$ operators and their compositions are all needed. For instance, $\mathcal{R}(AC) = CA$, $\mathcal{C}(AC) = TG$, $\mathcal{C}(\mathcal{R}(AC)) = GT$ and, of course, $\mathcal{R}(\mathcal{R}(AC)) = AC$. It is noteworthy that we have chosen the 2-word AC to generate the other 2-words, but this choice is arbitrary in the sense that any 2-word in the same equivalence class would do as well. Furthermore, observe that some equivalence classes have only two elements instead of four. This happens because the existence of fixed words whose definitions is as follows.

**Definition 6:** Given any operator $\mathcal{F}$ over $\mathcal{W}^k$, a fixed word of the operator $\mathcal{F}$ is defined as the k-word $w \in \mathcal{W}^k$ where $\mathcal{F}(w) = w$.

In other words, if we apply an operator over a word w and the result is the word w itself, then w is a fixed word of that operator. The definition does not guarantee the existence of fixed words, i. e., if they exist, then they most satisfy the definition. For instance, $\mathcal{C}$ operator has no fixed words, while $\mathcal{R}$ and $\mathcal{C} \circ \mathcal{R}$ operators do. The set $\{AT, TA, GC, CG\}$ is formed by the fixed words of the $\mathcal{C} \circ \mathcal{R}$ operator, since $\mathcal{C}(\mathcal{R}(AT)) = AT$, $\mathcal{C}(\mathcal{R}(TA)) = TA, \mathcal{C}(\mathcal{R}(GC)) = GC$ and $\mathcal{C}(\mathcal{R}(CG)) = CG$, and by the same token the set $\{AA, TT, CC, GG\}$ is formed by the fixed words of the $\mathcal{R}$ operator. Nevertheless, whenever k is an odd number the $\mathcal{C} \circ \mathcal{R}$ operator has no fixed words. The reason is that when k is an odd number, the complement of the letter in the center of the word is always a different letter as easily can be seen.

**Example 7:** If k=3, the 3-word set partition is

$$\mathcal{P}^3 = \left\{\begin{array}{l} \{AAA,TTT\},\{AAT,ATT,TAA,TTA\},\{TTG,CAA,AAC,GTT\},\{CTT,AAG,GAA,TTC\},\{ATA,TAT\}, \\ \{ATC,GAT,CTA,TAG\},\{ATG,CAT,GTA,TAC\},\{ACA,TGT\},\{TGA,TCA,AGT,ACT\},\{CCA,TGG,ACC,GGT\}, \\ \{GCA,TGC,ACG,CGT\},\{TCT,AGA\},\{GCT,AGC,TCG,CGA\},\{AGG,CCT,GGA,TCC\},\{CAC,GTG\}, \\ \{CAG,CTG,GAC,GTC\},\{CTC,GAG\},\{CCC,GGG\},\{GCC,GGC,CCG,CGG\},\{GCG,CGC\} \end{array}\right\}$$

The cardinality of $\mathcal{P}^k$ is given by $|\mathcal{P}^k| = 2^{k-1} + 4^{k-1}$, thus $|\mathcal{P}^1| = 2, |\mathcal{P}^2| = 6, |\mathcal{P}^3| = 20$ and so on. This number increases exponentially, for instance, $|\mathcal{P}^{10}| = 262656$.

## Generating Set

In example 6, we mentioned that one single k-word is sufficient to generate all other k-words in the same equivalence class. In other words, an equivalence class is fully represented by one single k-word. The collection of these representative k-words we will call Generating Set.

**Definition 7**: A subset containing one k-word of each equivalence class is called a Generating set.

**Example 8:** $\mathcal{G}^2 = \{AA, AT, CC, CG, AC, AG\}$ is a Generating set of $\mathcal{P}^2$ and so is $\mathcal{G}^{2\prime} = \{TT, TA, GG, GC, CA, GA\}$ or any set satisfying Definition 7.

**Example 9**:

$\mathcal{G}^3 = \begin{Bmatrix} AAA, AAT, TTG, CTT, ATA, ATC, ATG, ACA, TGA, CCA, \\ GCA, TCT, GCT, AGG, CAC, CAG, CTC, CCC, GCC, GCG \end{Bmatrix}$ is a Generating set of $\mathcal{P}^3$.

The concept of Generating set is a powerful one. Basically, we can represent all the words in $\mathcal{W}^k$ using just few words. For example, when k=3, there are $4^3 = 64$ words in $\mathcal{W}^3$, but we need only 20 words to represent $\mathcal{W}^3$. As k increases, the number of words in the Generating set tends to one quarter of the number of words in $\mathcal{W}^k$.

The generating set is not unique. There are no selection criteria of the generating set elements in the Definition 7. From a mathematical point of view, any element may represent its equivalence class; consequently the choice is arbitrary, and any selection criterion should be regarded as such. Although, mathematically, the Generating set elements choice is arbitrary, it does not mean that a Biology-based choice for the generating set elements does not exist.

## Generalized Chargaff's Second Parity Rule

Denoting the frequency operator by $\mathbb{F}$, then Chargaff's second parity rule may be written as $\mathbb{F}(A) \approx \mathbb{F}(T)$ and $\mathbb{F}(C) \approx \mathbb{F}(G)$. Using the $\mathcal{C}$ operator, this rule can be mathematically stated in a simplified form as $\mathbb{F}(w) \approx \mathbb{F}(\mathcal{C}(w))$, where w is any letter (nucleotide). Observe that Chargaff's rules is meant to be valid only to 1-words (mononucleotide), and it is naturally extended to k-word using the $\mathcal{C}$ and $\mathcal{R}$ operators as follows.

**Definition 8**: The Generalized Chargaff's Second Parity Rule, or simply GCSPR, if given by $\mathbb{F}(w) \approx \mathbb{F}(\mathcal{C}(\mathcal{R}(w)))$, where $w \in \mathcal{W}^k$.

The GCSPR is known in the literature as Symmetry Principle, and it states that the frequency of any k-word w is approximately equal to its complement reverse k-word frequency. Thus, in a sufficient long DNA sequence whenever w occurs, it is expected that its complement reverse k-word occurs as well. Although this property seems to be counterintuitive, it is expected just by chance in random DNA sequences provided that the mononucleotide frequencies follow CSPR. Actually, it is easy to show that in random DNA sequences that follows CSPR, the property is much stronger, i. e., in each equivalence class it is observed a complete equifrequency (Dong & Cuticchia, 2001): $\mathbb{F}(w) \approx \mathbb{F}(\mathcal{C}(\mathcal{R}(w))) \approx \mathbb{F}(\mathcal{R}(w)) \approx \mathbb{F}(\mathcal{C}(w))$. Thus, the interesting question about Symmetry Principle is why in actual DNA sequences the property is weaker, or in other words, why the symmetry is broken. It is clear that the reverse operator plays a major role in this case.

The equivalence class word frequencies are related to each other in the following way: whenever $\mathbb{F}(w) \approx \mathbb{F}(\mathcal{C}(\mathcal{R}(w)))$ holds so does $\mathbb{F}(\mathcal{C}(w)) \approx \mathbb{F}(\mathcal{R}(w))$ and vice versa as says the following theorem.

**Theorem 1**: Given $w \in \mathcal{W}^k$, $\mathbb{F}(w) \approx \mathbb{F}\big(\mathcal{C}(\mathcal{R}(w))\big) \Leftrightarrow \mathbb{F}(\mathcal{C}(w)) \approx \mathbb{F}(\mathcal{R}(w))$.

Theorem 1 proof is quite straightforward and uses the operators properties mentioned in this work.

The mathematical symbol $\Leftrightarrow$ means "if and only if". Thus Theorem 1 is quite strong, and it asserts that given an equivalence class, its word frequencies are grouped two by two, i. e., $\mathbb{F}(w) \approx \mathbb{F}\big(\mathcal{C}(\mathcal{R}(w))\big)$ and $\mathbb{F}(\mathcal{C}(w)) \approx \mathbb{F}(\mathcal{R}(w))$. In other words, Theorem 1 brings forward the very often unnoticed relationship between $\mathbb{F}(\mathcal{C}(w))$ $and$ $\mathbb{F}(\mathcal{R}(w))$. However, nothing is said about the relationship between those two groups' frequencies, i. e., their frequencies may be either different as in actual DNA sequences or similar as in random DNA sequences that follow CSPR.

As mentioned before, for every $k \geq 2$, $\mathcal{R}$ and $\mathcal{C} \circ \mathcal{R}$ operators may have fixed words. For instance, the $\mathcal{C} \circ \mathcal{R}$ operator fixed words satisfy GCSPR by definition. This is the case of w=CG, where $\mathcal{C}\big(\mathcal{R}(CG)\big) = CG$ and, of course, $\mathbb{F}(CG) = \mathbb{F}\big(\mathcal{C}(\mathcal{R}(CG))\big) = \mathbb{F}(CG)$. We have chosen this particular 2-word because it is known that usually in actual DNA sequences $\mathbb{F}(CG) \neq \mathbb{F}(GC)$, but there is no reason to expect otherwise since CG and GC are fixed words of the $\mathcal{C} \circ \mathcal{R}$ operator, therefore their frequencies should be equal to themselves, i. e., GCSPR is true by definition.

## *Self-Similarity*

There are $4^k$ k-words in $\mathcal{W}^k$, therefore from the very definition of frequency, we have

$$\sum_{i=1}^{4^k} \mathbb{F}(w_i) = 1. \qquad (1)$$

Equation 1 gives no insight of the k-word frequencies relationship. Each k-word $w_i$, where $i = 1, \ldots, 4^k$, seems to have its own independent frequency $\mathbb{F}(w_i)$. However, GCSPR implies that this is not the case. Within each equivalence class, as mentioned before, there are two frequency groups that as tidily related ($\mathbb{F}(w) \approx \mathbb{F}(\mathcal{C}(\mathcal{R}(w)))$ and $\mathbb{F}(\mathcal{C}(w)) \approx \mathbb{F}(\mathcal{R}(w))$). Using the definition of generating set, we can rewrite Equation 1 in a more convenient way.

Let $\mathcal{G}^k = \{g_1, g_2, \ldots, g_t\}$ be a Generating set of $\mathcal{P}^k$, where $t = |\mathcal{P}^k|$. By definition,

$$\sum_{i=1}^{t} \mathbb{F}(g_i) + \mathbb{F}(\mathcal{C}(\mathcal{R}(g_i))) + \mathbb{F}(\mathcal{C}(g_i)) + \mathbb{F}(\mathcal{R}(g_i)) = 1. \qquad (2)$$

In order to avoid an unnecessary complex notation, Equation 2 is actually a simplified version, since some equivalence classes have only two elements instead of four. In those cases, there are only two terms $\mathbb{F}(g_i)$ and $\mathbb{F}(\mathcal{C}(g_i))$ to sum up. It is Noteworthy that assuming GCSPR and Theorem 1, the independent variables in Equation 2 are $\mathbb{F}(g_i)$ and $\mathbb{F}(\mathcal{C}(g_i))$ for $i = 1, \ldots, t$. As mentioned before, whenever k is odd the $\mathcal{C} \circ \mathcal{R}$ operator has no fixed words, for instance k=3, implies that there are only 32 independents variables instead of 64 (total number of 3-words or codons).

The Equivalence Classes may be arranged in a matrix form where every row is a particular equivalence class; therefore this matrix has four columns: $g_i, \mathcal{C}(\mathcal{R}(g_i)), \mathcal{C}(g_i)$ and $\mathcal{R}(g_i)$. This matrix will be called Math Table just to emphasize that its definition is mathematical rather than biological in nature. Of course, the Math Table first column is the generating set.

**Example 10**: For k=3, we have the following Math Table

Table 1: Math Table for k=3. The Generating set is the first column.

| Class | $g_i$ | $\mathcal{C}(\mathcal{R}(g_i))$ | $\mathcal{C}(g_i)$ | $\mathcal{R}(g_i)$ |
|---|---|---|---|---|
| 1 | AAA | - | TTT | - |
| 2 | AAT | ATT | TTA | TAA |
| 3 | TTG | CAA | AAC | GTT |
| 4 | CTT | AAG | GAA | TTC |
| 5 | ATA | - | TAT | - |
| 6 | ATC | GAT | TAG | CTA |
| 7 | ATG | CAT | TAC | GTA |
| 8 | ACA | - | TGT | - |
| 9 | TGA | TCA | ACT | AGT |
| 10 | CCA | TGG | GGT | ACC |
| 11 | GCA | TGC | CGT | ACG |
| 12 | TCT | - | AGA | - |
| 13 | GCT | AGC | CGA | TCG |
| 14 | AGG | CCT | TCC | GGA |
| 15 | CAC | - | GTG | - |
| 16 | CAG | CTG | GTC | GAC |
| 17 | CTC | - | GAG | - |
| 18 | CCC | - | GGG | - |
| 19 | GCC | GCC | CGG | CCG |
| 20 | GCG | - | CGC | - |

The Math table number of rows increases exponentially, $2^{k-1} + 4^{k-1}$, for instance, when k=10 there are 262656 rows. Yet, organizing the Equivalence classes in this way, it is possible to recognize some hidden identities which otherwise would remain unnoticed. For instance, using real genomic data, no matter the Generating set elements choice, for the majority of the organisms, if we sum up the k-word frequencies in each column, no matter the value of k, we observed that

$$\sum_{i=1}^{t} \mathbb{F}(g_i) \approx \sum_{i=1}^{t} \mathbb{F}(\mathcal{C}(g_i)) \tag{3}$$

and

$$\sum_{i=1}^{t} \mathbb{F}(\mathcal{R}(g_i)) \approx \sum_{i=1}^{t} \mathbb{F}(\mathcal{C}(\mathcal{R}(g_i))) \tag{4}$$

Equations 3 and 4 cannot be derived from GCSPR, however if they are true, we may split Equation 2 into two summations as follows

$$\sum_{i=1}^{t} \mathbb{F}(g_i) + \mathbb{F}(\mathcal{C}(g_i)) + \sum_{i=1}^{t} \mathbb{F}(\mathcal{R}(g_i)) + \mathbb{F}(\mathcal{C}(\mathcal{R}(g_i))) = 1. \tag{5}$$

Now, using Equations 3 and 4, we may rewrite Equation 5 as

$$\sum_{i=1}^{t} 2\mathbb{F}(g_i) + \sum_{i=1}^{t} 2\mathbb{F}(\mathcal{R}(g_i)) \approx 1$$

or

$$\sum_{i=1}^{t} \mathbb{F}(g_i) + \sum_{i=1}^{t} \mathbb{F}(\mathcal{R}(g_i)) \approx \frac{1}{2}. \tag{6}$$

Analogously, it is possible to get

$$\sum_{i=1}^{t} \mathbb{F}(\mathcal{C}(g_i)) + \sum_{i=1}^{t} \mathbb{F}(\mathcal{C}(\mathcal{R}(g_i))) \approx \frac{1}{2}. \tag{7}$$

Equations 3, 4, 6 and 7 are supposed to be valid to every value of k, i. e., $k = 1, 2, ..., K^*$, where $K^*$ is an upper bound due the fact that actual DNA sequences are finite in length. Each value of k may be viewed as a different word scale, and no matter the scale the same invariant properties are preserved. The math table number of rows increases exponentially as mentioned before, yet Equations 3, 4, 6 and 7 still be valid no matter the number of rows (equivalence classes). This is a self-similarity property, and it is usually found in fractal-like objects. For this reason we call the new identities (equations 2,4, 6 and 7) fractal-like rules. Our former work (Yamagishi & Shimabukuro, 2008) is a particular case when k=1.

## *RESULTS*

The Math table concept is important because it helped us to discover hidden identities which otherwise would remain unnoticed. For instance, Equations 3 and 4 are only two examples of these hidden identities. Furthermore, using these identities and some simple mathematical reasoning, we were able to derive other two identities (Equations 6 and 7). In order to verify whether or not these derived identities were actually true, we have selected 33 organisms whose whole genomes were freely available at NCBI (http://www.ncbi.nlm.nih.gov/).

In the Figure 1, we show that 30 out of 32 genomes satisfy Equation 6, and consequently Equation 7, with great precision.

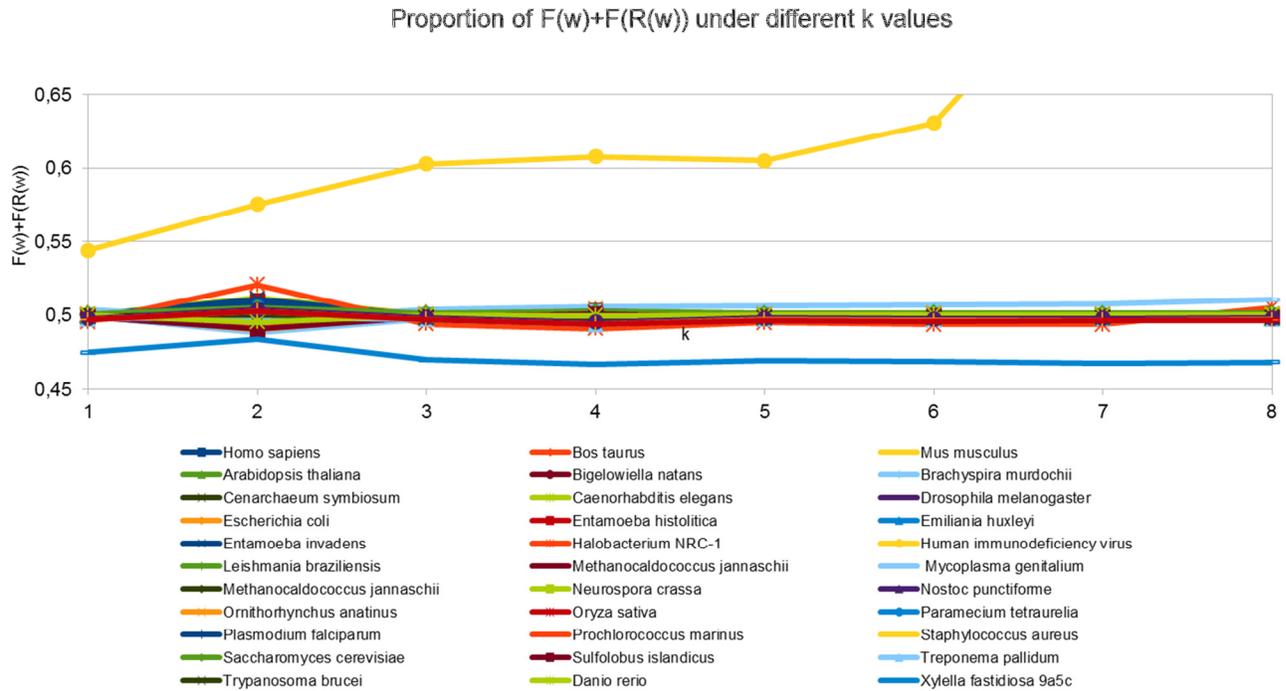

Figure 1: Equation 6 for 32 different organisms.

Figure 1 shows that only two organisms do not satisfy Equation 6: Human immunodeficiency virus (HIV) and *Xylella fastidiosa* 9a5c. HIV is actually a retrovirus whose genetic information is coded in a RNA molecule, and its sequence is shorter than those other organisms in Figure 1.

*Xylella fastidiosa* is a plant pathogenic bacteria, its first genome assembly was published back in 2000 (Simpson et al., 2000). Since then, other three alternative assemblies were deposited at NCBI: CP002165, M12 and M23. In Figure 2, we show the all four Xylella assemblies in regard Equation 6.

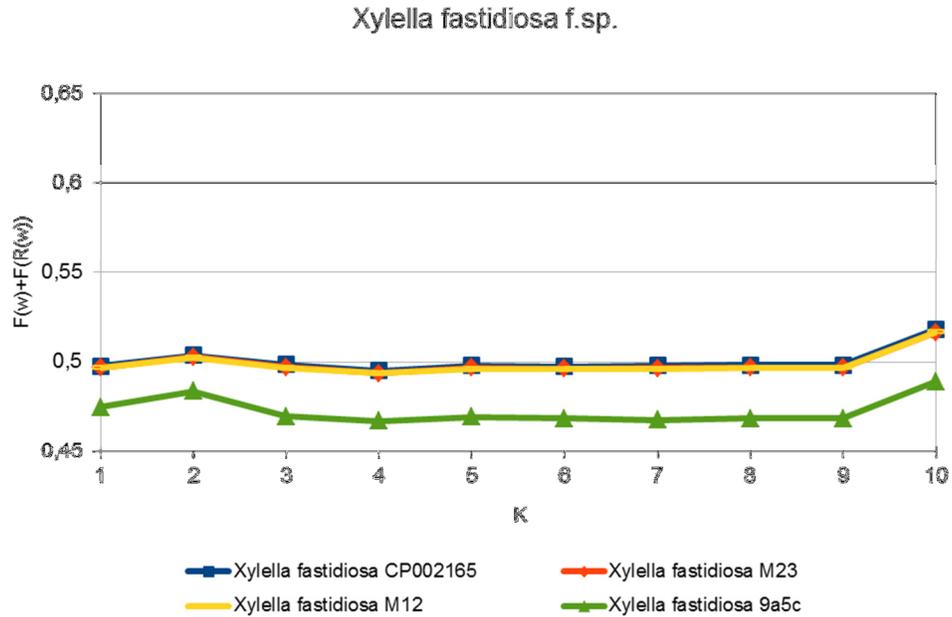

Figure 2: Only *Xylella fastidiosa* 9a5c significantly deviates from the expected value.

Interesting only the *Xylella fastidiosa* 9a5c assembly significantly deviates from the expected value. Actually, it does not satisfy Equations 3 and 4 either. As the other three assemblies do satisfy Equation 6, in order to better understand the assemblies' difference, we have performed a two-by-two alignment, and curiously only *Xylella fastidiosa* 9a5c presents several inversions compared to the other three alternative assemblies as is shown in Figure 3.

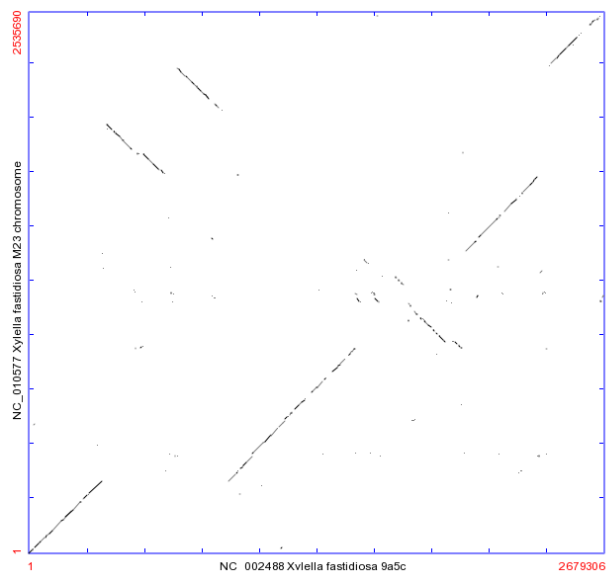

Figure 3: Xylella assemblies' comparison: observe that there are several sequence inversions. Source: Gmap at NCBI.

Considering that all tested bacterial genomic sequences satisfy Equation 6, it is necessary a deeper study to realize why the *Xylella fastidiosa* 9a5c assembly does not, while the other three Xylella's alternative assemblies do.

## *Random experiments and Generating sets*

Equations 6 and 7 split the Math Table in two sets with the same number of elements. For instance, for k=3, there are 64 words, and Equation 6 uses exactly 32 word frequencies, and so does Equation 7. Now, take any 64 frequencies whose sum is equal to one, then randomly choose 32 frequencies and sum them up. What is the expected value of the sum of those 32 randomly selected frequencies? The answer is given in Figure 4.

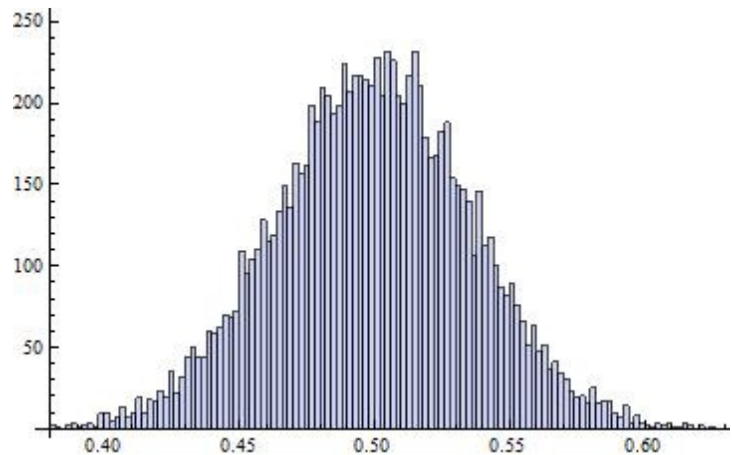

Figure 4: Histogram of 10.000 randomly selected 32 frequencies sum.

As shown in Figure 4, the expected value is 0.5. Thus, is the Math Table arrangement important? Is Equation 6 predicted value expected just by chance? In order to answer these questions, beyond this completely random data and random selection case, there are other two cases to consider: (i) real data, but random selection and (ii) real data, but random permutation within each Equivalence class, i. e., different Generating sets (does the result depend on the Generating set elements choice?).

Now to assess the first case, we have selected the Xylella M23 frequencies for k=3, and 10.000 times randomly selected 32 frequencies (without take in account the equivalence classes). The histogram is illustrated in Figure 5.

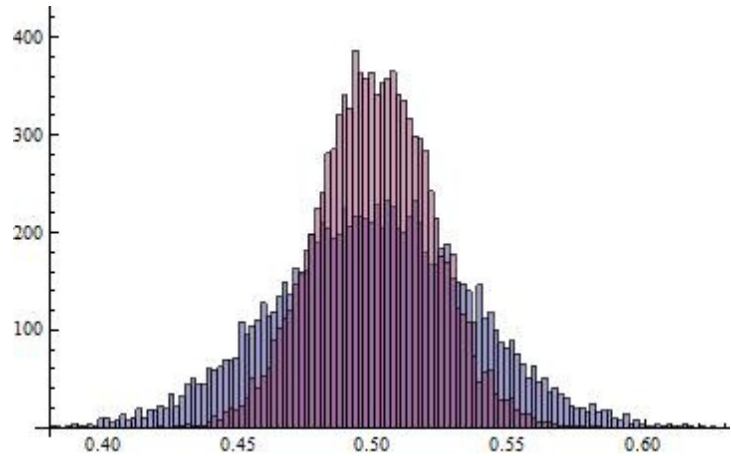

Figure 5: In pink the real data, but randomly selected. As expected the histogram is more concentrated than the totally random case (in blue).

The narrower histogram corresponds to the real data. As real data preserves GCSPR, it is expected just by chance that some randomly selected frequencies resemble the arrangement of the math table.

Finally the second case with real data, but the selection is actually a random permutation within each equivalence class (Figure 6) which means that we are randomly changing the Generating sets.

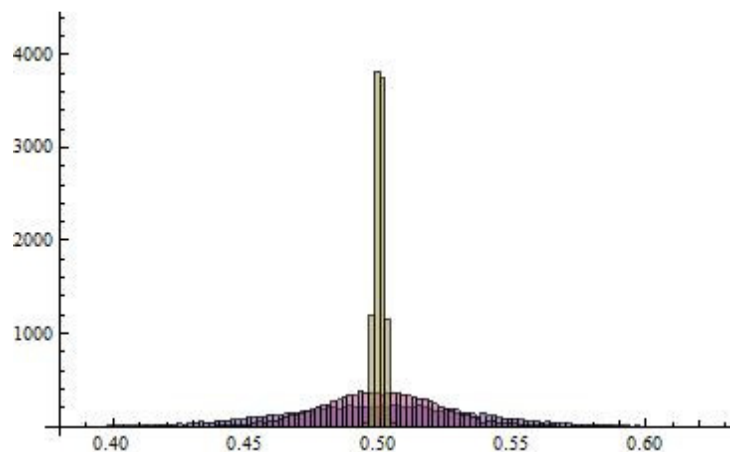

Figure 6: The higher peak corresponds to the random permutation within each equivalence class.

Although it is not possible to see in Figure 6, the real data range of the values goes from 0.496 to 0.506 which is almost 0.5. This last case shows that the Math Table is really important and the validity of Equations 6 and 7 do not depend on the Generating set choice. Furthermore, if we use genomes whose assembly qualities are higher, then the error appears in the 4<sup>th</sup> decimal digit as shown in Figure 7 which suggests that the better the genome assembly quality, closer to the expected value it will be.

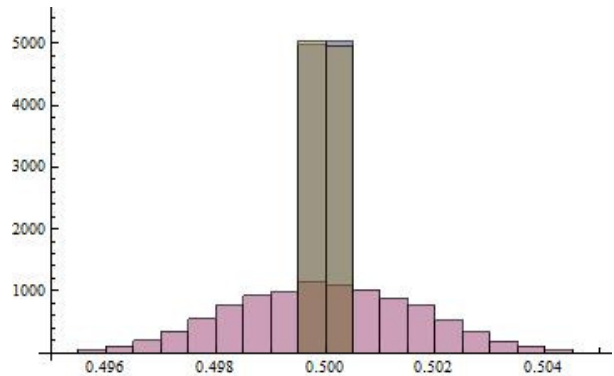

Figure 7: High quality genome assemblies (Arabidopsis thaliana and Homo sapiens (blue)) compared to Xylella M23 (pink) genome assembly.

## *CONCLUSION*

We have enriched Chargaff's grammar of Biology with four new identities. Two of them were revealed after the Equivalence classes' organization into the Math Table. The other two were derived from the first two. It was shown that these rules are valid for a large set of organisms: bacteria, plants, insects, fish and mammals. It is noteworthy that no matter the word length the same pattern is observed (self-similarity). To the best of our knowledge, this is the first invariant genomic properties publish so far, and in Science invariant properties are invaluable ones and usually they have practical implications.

From our point of view, our work has an aesthetical value, as long as it shows some hidden regularities (invariant) across the genome of several species. We hope that it may be of practical importance as well. For instance, recently, the new high-throughput DNA sequencing technologies have produced a huge amount of short read data, and it is known (Hansen et al., 2010) that these data might be biased for unknown reasons. In some studies, it is necessary a reference-free validation of short read data, and novel methodologies have already been proposed (Schroder et al., 2011) to address this problem. The basic idea is to measure some invariant properties to assess the data quality. If the measured properties show variance, then the data might be biased. The new frequency rules presented in this work may perfectly be used for this purpose.

We started this work quoting Chargaff, and we would like to end in the same way: "Einstein is somewhere quoted as having said: `the ununderstandable about nature is that it is understandable'. I think, he should have said: `that it is explainable'. These are two very different things, for we understand very little about nature." (Chargaff, 1975).

## *Acknowledgements*

We dedicate this paper to the memory of the great Scientist *Erwin Chargaff* and to the memory of Michel's dear mother *Marlene M. F. Belleza* who passed away this year. We thank *Donald Forsdyke* for in depth discussions. This work was performed with support of the Applied Bioinformatics Laboratory at EMBRAPA.

## *REFERENCES*

Avery, O. T., MacLeod, C. M., McCarty, M. (1944) Studies on the chemical nature of the substance inducing transformation of pneumococal types, J. Exp. Med., 79, 137

Bell S. J. and Forsdyke, D. R. (1999) Deviations from Chargaff's Second Parity Rule Correlate with Direction of Transcription, J. Theo. Bio., 197, 63-76

Chargaff, E. (1950) Chemical specificity of nucleic acids and mechanism of their enzymatic degradation, Experimentia, 6, 201

Chargaff, E. (1971) Preface to a Grammar of Biology: A hundred years of nucleic acid research, Science, 172, 637-642

Chargaff, E. (1975) A fever of reason, Annu. Rev. Biochem, 44:1-20

Dong, Q. Cuticchia, A. J. (2001) Compositional symmetries in complete genomes, Bioinformatics, 17, 557-559;

Forsdyke, D. R. and Bell, S. J. (2004) A discussion of the application of elementary principles to early chemical observations}, Applied Bioinformatics, 3, 3-8

Hansen, K. D., Brenner, S. E. and Dudoit, S. (2010) Biased in Illumina transcriptome sequencing caused by random hexamer priming, NAR, 38, e131

Kong S-G, Fan W-L, Chen H-D, Hsu Z-T, Zhou N, et al. (2009) Inverse Symmetry in Complete Genomes and Whole-Genome Inverse Duplication, PLoS ONE 4(11): e7553. doi:10.1371/journal.pone.0007553

Mitchell, D. and Bridge, R. (2006) A test of Chargaff's second rule, BBRC, 340, 90-94

Prabhu, V. V. (1993) Symmetry observation in long nucleotide sequences, Nucleic Acids Res., 21, 2797-2800

Schroder, J., Balley, J., Conway, T. and Zobel, J. (2011)  Reference-Free Validation of Short Read Data, PLoS One, 5, e:12681

Simpson, A. J. G. et al. (2000) The genome sequence of the plant pathogen Xylella fastidiosa, Nature, 406, 151-157

Yamagishi, M. E. B. and Shimabukuro, A. I. (2008) Nucleotide Frequencies in Human Genome and Fibonacci Numbers, Bull. Math. Biology, 70, 643-653

Watson, J. D. and Crick, F. H. C. (1953) Molecular Structure of Nucleic Acids, Nature, 4356, 737